\documentclass[twocolumn]{aastex631}
\usepackage[a4paper,marginpar=10pt,headheight=20pt, right=1cm, top=1.5in, left=1in, bottom=0pt, footskip=0pt, footnotesep=10pt]{geometry}
\usepackage[utf8]{inputenc}
\usepackage[T1]{fontenc}
\usepackage[toc,page]{appendix}

\usepackage{amsfonts}
\usepackage{amsmath}
\usepackage{amssymb}
\usepackage{nicefrac}
\usepackage{bm}

\usepackage{soul}
\usepackage{xspace}
\xspaceaddexceptions{]\}} % https://tex.stackexchange.com/questions/15252/why-does-xspace-behave-differently-for-parenthesis-vs-braces-brackets/15254#15254

\usepackage{braket}

\usepackage{enumerate}
\usepackage{enumitem}
\usepackage{array}
\usepackage{tabularx}
\usepackage{floatrow}

\usepackage{graphicx,array}
\usepackage[caption=false]{subfig}
\usepackage{epsfig}
\usepackage{longtable}

\usepackage{natbib}
\setcitestyle{round}

\usepackage{hyperref}

\newcommand{\D}[2][]{
  \ensuremath{\frac{d {#1}}{d {#2}}}}%

\newcommand{\Lvec}{{\ensuremath{\bm{{L}}}}\xspace}%
\newcommand{\lvec}{{\ensuremath{\bm{{\ell}}}}\xspace}%
\newcommand{\lmax}{\ensuremath{\ell_\textnormal{max}}\xspace}%
\newcommand{\dz}{\ensuremath{\Delta z}\xspace}%
\newcommand{\degsq}{\ensuremath{\textnormal{deg}^2}\xspace}%
\newcommand{\ngal}{\ensuremath{n}\xspace}%
\newcommand{\dndz}{\ensuremath{\D[n]{z}}\xspace}%
\newcommand{\diffSNRLz}{\ensuremath{\textnormal{SNR}_\Delta(L,z)}\xspace}

\newcolumntype{C}{>{\centering\arraybackslash}X}%
\newcolumntype{R}{>{\raggedleft\arraybackslash}X}%
\newcolumntype{L}{>{\raggedright\arraybackslash}X}%

\def\aap{{ A\&A}}

\def\apjl{{ApJL}}

\def\apjs{{ApJS}}

\begin{document}

\title{Gravitational lensing of galaxy clustering}
\author{Brandon Buncher}
\affiliation{Department of Physics, University of Illinois Urbana-Champaign}
\author{Gilbert Holder}
\affiliation{Department of Physics, University of Illinois Urbana-Champaign}
\author{Selim C. Hotinli}
\affiliation{Perimeter Institute for Theoretical Physics}
\date{\today}

\begin{abstract}
  We investigate lensing reconstruction using the clustered galaxy distribution as a source field, using both the traditional cosmic microwave background quadratic estimator and a shear-only estimator. 
  We calculate the expected signal-to-noise ratio of the cross power spectrum of such reconstructions with cosmic shear measurements for an LSST-like galaxy survey. Modeling the galaxy field as a Gaussian random field, we find that there is substantial clustering signal in the source field at angular scales substantially smaller than those typically used by CMB reconstructions. 
  The expected signal-to-noise for cross-correlations in LSST from cosmic shear is $\sim$60 in the presence of shape noise, while cross-correlating with a sample-variance limited mass map would have signal-to-noise in the hundreds. This type of cross-correlation could be used as a way to identify systematic errors in lensing studies and is just one example of many possible higher order correlations in galaxy surveys that may contain substantial cosmological information. 
\end{abstract}

\section{Introduction}

There is large scale structure (LSS) in the universe, as evidenced by the spatial clustering of galaxies \citep{PeeblesLSS}. This LSS deflects light as it propagates through the inhomogeneous universe \citep{BartSch}. These deflections can occasionally lead to spectacular Einstein rings and multiple images, but usually will lead to only small distortions to the shapes and brightnesses of background objects. This has been measured in galaxy surveys through the changes in the detailed shapes of galaxies \citep{SDSSLens2,DESLens}, and has been measured in the distortion to the characteristic clustering patterns in the cosmic microwave background \citep{PlanckLens2018,SPT,SPTLensing,ACT,ACTLensing,POLARBEAR,BICEPKeck}.

In this paper we investigate the gravitational lensing of the clustering of background galaxies by foreground matter, analogous to the case of gravitational lensing of the CMB. A related method was recently discussed in detail by \citet{MBiasLensing} using an estimator that combined this information with information about the magnification bias to increase the signal-to-noise ratio (SNR). We determine the viability of a different (shear-based) estimator and investigate the prospects for measuring this effect in cross-correlation with cosmic shear measurements. 

In the case of CMB lensing, the workhorse tool that is widely used is a quadratic estimator \citep{QEHu}, which amounts to multiplying together two CMB maps that each have their own filtering. The filtering of the maps is chosen to create an unbiased, minimum variance estimator, which optimizes the SNR of the resulting lens reconstruction. If this lens reconstruction is then correlated with a galaxy-based LSS tracer, the result could be expressed as a 3-pt function, as it involves products of two CMB maps and an LSS map. As we will show below, for the similar case of lensing of the galaxy clustering distribution by foreground matter, the reconstruction of the foreground lensing potential will involve two versions of the background galaxy number count field, and then cross-correlation with an LSS tracer such as cosmic shear can be expressed as a galaxy-galaxy-shear 3-pt function. If we instead used the foreground galaxy counts, then there is a 3-pt function that would be quadratic in the background galaxy number counts and linear in foreground galaxy number counts. In general, the three fields going into these estimates can be from diverse LSS tracers: the background 2-pt function could be cross-correlations between different LSS tracers, and the foreground could be taken to be any LSS tracer of the foreground matter \citep{GeneralQE}. In this work, we only look at the simplest case of using the auto-correlation of galaxies in the background being lensed by foreground matter in a way that leads to a measured shear of the same background material, mainly to keep the accounting simple.

This work is closely related to other work that has looked at the lensing of the clustering of LSS tracers. Prior work on lensing reconstruction has primarily used the CMB as a light source \citep{QEHu}. However, modifications of the quadratic estimator have been applied to non-CMB sources, particularly spectral line intensity maps (LIM). The closest works to ours use the standard quadratic estimator to perform lensing reconstructions from measurements of the Lyman-$\alpha$ forest \citep{MetcalfLyAlph}, the cosmic infrared background (CIB) \citep{SchaanCIB,ForemanLIM}, and 21cm lines \citep{Pen21,ZahnZal21,LuPen21a,Torres21,Port212014,Port212015,FullSky21}; other 21cm reconstruction methods include \citep{LuPen21b,Jalilvand21,ZhangPen21a,PregalHI2007a,PregalHI2007b,PregalHIInfo}.

An important point of focus in each of these studies has been the non-Gaussianity of the background field being lensed. It has been shown \citep{SchaanCIB,ForemanLIM,FengCIB} that there can be substantial biases in the recovered lensing power spectrum if the non-Gaussianity is ignored. Nevertheless, for this preliminary investigation we will disregard the non-Gaussianity of the field and its uncertainties in what follows, so the results presented below will no doubt require future work to study these effects. These corrections arise because non-linear gravitational collapse leads to an intrinsic background of spatial variation in the statistics of small-scale clustering, the magnitude of which depends on factors including the spatial scales probed, redshift, and survey parameters \citep{ForemanLIM}. However, some lensing estimators that are optimal for a Gaussian source field have been shown to be resistant to non-Gaussian noise. \citet{ForemanLIM} constructed a bias-hardened estimator that included contributions from the density field trispectrum that corrected for mild gravitational nonlinearities, while \citet{LuPen21b} calibrated their estimator with N-body simulations to create a set of lensing weights that are optimal for non-Gaussian fields.

In this work, we closely follow the procedures described by \citet{LensQuEst}, who present a method for lensing reconstruction that is less affected by foreground contamination and non-Gaussianities. In \citet{LensQuEst}, the authors construct a convergence field hybrid quadratic estimator for a Gaussian CMB source field. The hybrid estimator is comprised of an inverse-variance weighted linear combination of a magnification estimator and a shear estimator. It used the shear-only component for small scales and both the convergence and shear components for large scales. The authors found that the hybrid estimator outperformed the SNR of the standard quadratic estimator and that the shear component was less affected by foreground contamination \citep{LensQuEst}.
 \citet{QCSShear} generalized this estimator for full-sky reconstruction and verified using simulations that these estimators are less affected by foreground contamination than the standard QE.
 
 {The way that foregrounds contaminate CMB lensing measurements is through their non-Gaussianity; reducing foreground contamination means reducing the impact of the non-Gaussianity induced by non-linear structure formation. It is therefore expected that these new estimators should also be less sensitive to gravitationally induced non-Gaussianity of the background source field.  \citet{SchaanCIB} conjectured that a shear-only estimator would be robust to non-Gaussian contamination for CIB fields, which consist of a superposition of point sources (galaxies) that trace the nonlinear density field. Furthermore, in what follows below we will primarily focus on cross-correlation, where the impact of source field non-Gaussianity may also be lower.}

\citet{LuPen21a} present a similar hybrid estimator for a 21 cm source field. The authors combined the magnification and shear estimators at all spatial scales. They show that, for a Gaussian source field, their estimator is equivalent to the 3D optimal quadratic deflection estimator introduced by \citet{ZahnZal21}, but was less affected by gravitational nonlinearities when reconstructions were performed using a non-Gaussian source field. Like \citet{LensQuEst}, the authors found that the hybrid estimators were also less susceptible to this foreground contamination.

{The estimator we use \citep{LensQuEst} was developed for CMB lensing reconstruction, which uses a Gaussian source field. We propose here to perform reconstructions using a galaxy number count source field, which is highly non-Gaussian, and detailed simulations will ultimately be required to assess the robustness of the shear estimator for this application. The 21 cm source field, with large fluctuations in ionization in addition to the density fluctuations, is expected to be substantially more non-Gaussian than a typical galaxy distribution, so the results of \citet{LuPen21a} are promising in finding that the shear estimator is not substantially biased by this non-Gaussianity.}

{Typical CMB lensing techniques probe large spatial scales ($\ell \lessapprox 3,\!000$), but we found that most of the SNR for the cross-correlation between the observed and theoretical power spectra is contained at very large $\ell$ (up to $\ell \approx 70,\!000$). These are similar scales to what was explored in both \citet{LuPen21a} and \citet{SchaanCIB}, where there were no obvious problems with applying these techniques other than the non-Gaussianity of the field leading to the traditional weights becoming sub-optimal.}

We have found one other work that uses galaxy number counts as a source field for weak lensing reconstruction in a similar manner. In \citep{MBiasLensing}, the authors demonstrate that measurements of the magnification bias can be combined with the information from the background galaxy density field to reconstruct the convergence power spectrum using a quadratic estimator at a high SNR. The magnification bias has been well-studied both through measurements and forecasts \citep{Scranton2005,MBiasMeas1,MBiasMeas2,MBiasMeas3,FluxBias}; however, for the purposes of this work, we will ignore its effects in order to focus on the particular information that is carried in the distortion of the clustering field.
The effects of galaxy clustering and magnification bias on weak lensing reconstruction were also examined in \citep{WLMagPhot}. Noise from intrinsic galaxy clustering is much larger than the lensing signal, so the authors developed a multi-band linear estimator for the convergence field that minimizes the noise introduced by intrinsic clustering. While this work does not use galaxy number counts as a source field, it supports the conclusions of \citep{MBiasLensing} that magnification bias provides valuable information for weak lensing reconstruction.

We perform projections of the SNR for a survey with parameters consistent with the Legacy Survey of Space and Time at the Vera C. Rubin Observatory (LSST) \citep{LSSTOverview}. To estimate the expected properties of LSST galaxies we will use the CosmoDC2 simulations \cite{CosmoDC2}; for consistency, in our calculations we will also use the cosmology assumed for that simulation. This parameter set is similar to those of the best fit WMAP-7 parameter set \citep{WMAP7}: $\omega_\textnormal{cdm}=0.1109$, $\omega_\textnormal{b}=0.02258$, $n_s=0.963$, $h=0.71$, $\sigma_8=0.8$, and $w=-1.0$. We investigate $r$-band magnitude cuts of $m_r<23$ and $m_r<25$, chosen to correspond roughly with the Dark Energy Survey (DES) limiting magnitude \citep{DESGalLens} and the 5$\sigma$ single visit limiting magnitude of the LSST survey \citep{LSSTHandbook}, respectively. The LSST 5$\sigma$ 10 year coadd. dataset would increase the limiting magnitude to around $m_r < 27$. However, we find that information from very faint galaxies was not needed to obtain a high SNR using our method. We expect that the full LSST 5$\sigma$ 10 year coadd. dataset would increase the SNR.

The next section lays out the theoretical framework for our calculations, and the following section describes the computational setup. We then present the expected SNR for an LSST-like survey and a DES-like survey, and then close with a discussion.

\section{Theory}

In this section we describe the theoretical framework, including a brief overview of gravitational lensing, a discussion of lensing estimation using quadratic estimators, and a method for calculating the expected lensing power spectrum for a screen of sources at a given redshift. 

\subsection{Lensing basics}

We can observe the galaxy number density $\ngal$ at redshift $z$ and locations $\vec{\theta}$ to be

\begin{equation}
    \ngal(\vec{\theta})= \ngal^u(\vec{\theta}+\vec{\alpha}),
\end{equation}
where $\vec{\alpha}$ is the gravitational deflection and the superscript $u$ indicates the field that would be observed in the absence of gravitational lensing (the unlensed field). We here ignore the effects of magnification bias, where selection effects for the galaxy sample are affected by the lensing magnification.

In the weak lensing regime (where the deflection $\alpha$ is small), the lensed number density can be approximated via a Taylor series:

\begin{equation}
    \ngal(\vec{\theta}) \approx \ngal^u(\vec{\theta})+\nabla \ngal^u(\vec{\theta}) \cdot \vec{\alpha}. \label{eq:DeflExp}
\end{equation}

The deflection angle can be expressed as the gradient of a deflection potential, $\vec{\alpha}=\nabla \psi$. The deflection potential $\psi$ is the line-of-sight projection of the Newtonian gravitational potential $\Phi$ \citep{BartSch}:

\begin{align}
    \psi=\frac{2}{c^2}\frac{D_\textnormal{ds}}{D_\textnormal{d}D_\textnormal{s}}\int \Phi\left(D_\textnormal{d}\,\vec{\theta}, z\right)dz, \nonumber
\end{align}
where $D_\textnormal{d}$ is the distance between the observer and the lens, $D_\textnormal{s}$ is the distance between the observer and the source, and $D_\textnormal{ds}$ is the distance between the source and the lens.

The effect of weak lensing is either to expand or contract the apparent solid angle of a galaxy or cluster, known as ``convergence,'' or to systematically stretch or squeeze them in orthogonal directions, known as ``shear.''  This affects the perceived clustering of galaxies, leading to alterations to the observed galaxy power spectrum. The convergence field $\kappa$ and two components of the shear field $\gamma_+$ and $\gamma_\times$ can be expressed as a transformation of the deflection potential $\psi$:

\begin{align}
    &\kappa=-\frac{1}{2}\left(\partial_x^2+\partial_y^2\right)\psi=-\frac{1}{2}\nabla^2\psi, \nonumber\\
    &\gamma_+=-\frac{1}{2}\left(\partial_x^2 - \partial_y^2\right)\psi, \\
    &\gamma_\times=-\partial_x\partial_y\psi \nonumber
\end{align}
for some orthogonal angular directions $\hat{x}$ and $\hat{y}$ where here we ignore sky curvature and approximate the sky as being a Cartesian grid. 

A value of $\kappa > 0$ represents image dilation (an increase in solid angle of the object), while $\kappa < 0$ represents contraction. A value of $\gamma_+ > 0$ indicates stretching in the horizontal ($\hat{x}$) direction, while $\gamma_+ < 0$ represents stretching in the vertical ($\hat{y}$) direction. A value of $\gamma_\times > 0$ represents stretching along the line $y=-x$, while $\gamma_\times < 0$ represents stretching along the line $y=x$. Because each of the lensing observables are constructed from derivatives of the deflection potential, this means that they may be reconstructed from one another.

\subsection{Lensing reconstruction with quadratic estimators}

The foreground lensing deflection can be reconstructed using a ``quadratic estimator'';  multiplying Eq. \eqref{eq:DeflExp} by $\nabla \ngal^u(\vec{\theta})$ leaves the second term proportional to the square of the galaxy density field gradient multiplied by the deflection \citep{QEHu}. This is known as a quadratic estimator because it uses the square of the observed field to reconstruct the lensing deflection. An optimal quadratic estimator (i.e. a quadratic estimator that minimizes the noise in the reconstructed field) can be constructed that takes into account the noise in the background field; a derivation of this estimator can be found in the appendix of many papers, including in \citet{LensQuEst}, which we follow closely in this work.

On large scales, the main effect of the lensing distortion is to simply change the observed small-scale clustering pattern of the background field. The lensing effect therefore breaks statistical isotropy: the small-scale power spectrum of the background field will be observed to vary from place to place on the sky because of the variations in the large-scale lensing effects.

In the limit of large-scale lensing of small-scale structure, the optimal quadratic estimator can be shown to be a mix of a convergence estimator and a shear estimator \citep{Bucher1}. On small scales, this separation between the estimators breaks down, but it has been shown that the shear estimator is more robust to non-Gaussian foregrounds and experimental non-idealities \citep{LuPen21a,LensQuEst}.

The lensing maps are reconstructed using a quadratic estimator that uses two combinations of the background galaxy density field along with a Fourier space lensing kernel $g_{\Lvec, \lvec}$ that captures the lensing-related correlations. We use a notation where $\Lvec$ refers to an angular mode in the foreground lensing potential, while $\lvec$ refers to angular modes in the background source that is being lensed.

The standard quadratic estimator is a minimum variance quadratic estimator that is commonly used for reconstruction of the CMB lensing deflection potential $\psi$. The QE is calculated using the Fourier transform (spherical harmonics decomposition) of the background field. Throughout this paper, we use the convention that \lvec indexes the multipole of the Fourier transform of the background field and \Lvec indexes the multipole of the Fourier transform of the lensing field. Traditionally, the standard quadratic estimator uses the background CMB temperature field $\Theta$ and its power spectrum $C_\ell^\Theta$ to calculate the deflection field; in our case, we replace the observed temperature field $\hat{\Theta}$ with the observed galaxy number density $n$. The standard quadratic estimator performs the reconstruction by calculating correlations between the observed number density field at different locations $\ngal_\lvec$ and $\ngal_{\lvec-\Lvec}$; this convolution is weighted by the lensing kernel $g_{\lvec,\Lvec}$. The reconstruction of the observed lensing deflection potential in Fourier space $\psi(\Lvec)$ using the galaxy number density field is calculated \citep{LewChal}

\begin{align}
    \psi(\Lvec) = N_\Lvec\int\frac{d^2\lvec}{2\pi}\ngal_\lvec\ngal^*_{\lvec-\Lvec}g^\psi_{\lvec, \Lvec}. \label{eq:QE}
\end{align}

Here, the reconstruction noise $N_\Lvec$, which is the variance of the QE, is defined as \citep{LewChal}

\begin{align}
    \frac{1}{N_\Lvec^\psi} = \int\frac{d^2\lvec}{(2\pi)^2}\left[\left(\Lvec-\lvec\right)\cdot\Lvec C_{\left|\lvec-\Lvec\right|}^0 + \lvec\cdot\Lvec C_\ell^0\right]g_{\lvec, \Lvec}^\psi, \label{eq:NLQE}
\end{align}

where $C_\lvec^0$ is the power spectrum of $n$ and $C_{|\lvec - \Lvec|}^0$ is the power spectrum of the shifted number density field. The lensing kernel $g_{\lvec, \Lvec}$ for the lensing potential is defined as \citep{LewChal}

\begin{align}
    g_{\lvec, \Lvec}^\psi = \frac{(\Lvec-\lvec)\cdot\Lvec C_{\left|\lvec-\Lvec\right|}^0 + \lvec\cdot\Lvec C_\ell^0}{2C_\ell^\textnormal{tot}C_{\left|\lvec-\Lvec\right|}^\textnormal{tot}} \label{eq:gQE}
\end{align}
where $C_\lvec^0$ is the power spectrum of the background field being lensed and $C_\ell^\textnormal{tot}$ is the same power spectrum plus noise. For CMB lensing, this noise will be a combination of detector noise and foregrounds, while for our purposes, this noise will be the shot noise of the background field.

At large scales, the standard quadratic estimator can be broken down into a combination of a magnification-only estimator and shear-only estimator. The estimator is broken down by dividing the standard quadratic estimator weight function into the contributions from the monopole ($m=0$) $g^0_{\Lvec, \lvec}$ (the contribution from convergence) and quadrupole  ($m=2$) $g^2_{\Lvec, \lvec}$ (the contribution from shear). Higher order terms are neglected in this work. 

The lensing kernels for the convergence field are defined as \citep{Bucher1,LensQuEst}

\begin{align}
    &g_{\Lvec, \lvec}^{\kappa,0}=\frac{C_\ell^0}{2\left(C_\ell^\textnormal{tot}\right)^2}\frac{\textnormal{d}\ln\ell^2C_\ell^0}{\textnormal{d}\ln\ell} \label{eq:Mkern}\\
    &g_{\Lvec, \lvec}^{\kappa,2}=\cos\left(2\theta_{\Lvec, \lvec}\right)\frac{C_\ell^0}{2\left(C_\ell^\textnormal{tot}\right)^2} \frac{\textnormal{d}\ln C_\ell^0}{\textnormal{d}\ln\ell}. \label{eq:Skern}
\end{align}
In our case, we will model the noise as being from the shot noise in the galaxy number counts. In principle, this could be considered to be part of the signal that would contribute to the convergence estimator (Eq. \eqref{eq:Mkern}).

The magnification ($m\,=\,0$) and shear ($m\,=\,2$) estimators of the convergence field are defined by \citep{LensQuEst}

\begin{align}
    \hat{\kappa}^m_\Lvec=\frac{\int\frac{d^2\lvec}{(2\pi)^2}g^{\kappa,m}_{\Lvec, \lvec}n_\lvec n_{\Lvec - \lvec}}{\frac{2\Lvec}{L^2}\int\frac{d^2\lvec}{(2\pi)^2}g^{\kappa,m}_{\Lvec, \lvec}\left[\lvec C_\ell^0+(\Lvec - \lvec)C_{\left|\Lvec - \lvec\right|}^0\right]}, \label{eq:kappamhat}
\end{align}
where $n_\lvec$ is the Fourier transform of the galaxy number density field. The reconstruction noise $N_\Lvec^m$ is the variance of each estimator:

\begin{align}
    N_\Lvec^{\kappa,m}=\frac{\int\frac{d^2\lvec}{(2\pi)^2}g^m_{\Lvec, \lvec}\left(g^{\kappa,m}_{\Lvec, \lvec}+g^{\kappa,m}_{\Lvec, \Lvec - \lvec}\right)C_\ell^\textnormal{tot}C_{\left|\Lvec - \lvec\right|}^\textnormal{tot}}{\left\{\frac{2\Lvec}{L^2}\int\frac{d^2\lvec}{(2\pi)^2}g^{\kappa,m}_{\Lvec, \lvec}\left[\lvec C_\ell^0+\left(\Lvec - \lvec\right)C_{\left|\Lvec - \lvec\right|}^0\right]\right\}^2}, \label{eq:NL}
\end{align}
where again the $m\,=\,0$ mode is the magnification noise and the $m\,=\,2$ mode is the shape noise.

We will be comparing our lensing reconstructions using the lensing of the galaxy correlation function to what can be obtained from a cosmic shear survey that measures the shapes of individual galaxies as a proxy for the gravitational shear.
We thus consider the contribution of shape noise, $N_\gamma$, to lensing reconstructions based on cosmic shear measurements; $N_\gamma$ represents the intrinsic variance of the galaxy ellipticities that act as noise in the shear measurements for each galaxy in a cosmic shear survey. As is specified in \citet{LSSTHandbook}, we take shape noise to be

\begin{align} \label{eqn:ShapeNoise}
    N_\gamma \equiv \frac{\gamma_\textnormal{rms}}{\sqrt{\ngal}},
\end{align}
where $\gamma_\textnormal{rms}=0.28$ is the shape noise per galaxy expected for an LSST-like survey and \ngal is the average number of galaxies per steradian.

\subsection{The angular power spectra of LSST galaxy subsamples}

The quadratic estimators we use require the angular power spectra of each galaxy subsample, $C_\ell$. In a real survey these power spectra will be measured in the data, but for this forecasting study we need to calculate the expected power spectra. We use CosmoDC2 mock galaxy catalogs \citep{CosmoDC2} to guide the inputs for the calculation. The calculation is performed using the Limber equation \citep{LimberEq} assuming a basic linear bias model:

\begin{align}
    C_\ell(z)=\frac{H_0^2}{c^2}\int\frac{1}{n^2}\left(\dndz\right)^2b^2(z) P_\delta\left(\frac{\ell}{\chi}, z\right)\,\frac{d\chi}{\chi^2}, \label{eq:Limber}
\end{align}
where $\chi$ is the comoving distance, $H_0$ is Hubble's constant, $\dndz$ is the galaxy density distribution, $b(z)$ is the linear galaxy bias, which represents the ratio of the mean galaxy overdensity to the mean dark matter overdensity, and $P_\delta\left(k, z\right)$ is the total matter power spectrum at wavenumber $k$ and redshift $z$. 

\begin{figure*}
    \begin{minipage}{\textwidth}
    \centering
        \subfloat[\centering \dndz]{
        \includegraphics[width=0.33\textwidth]{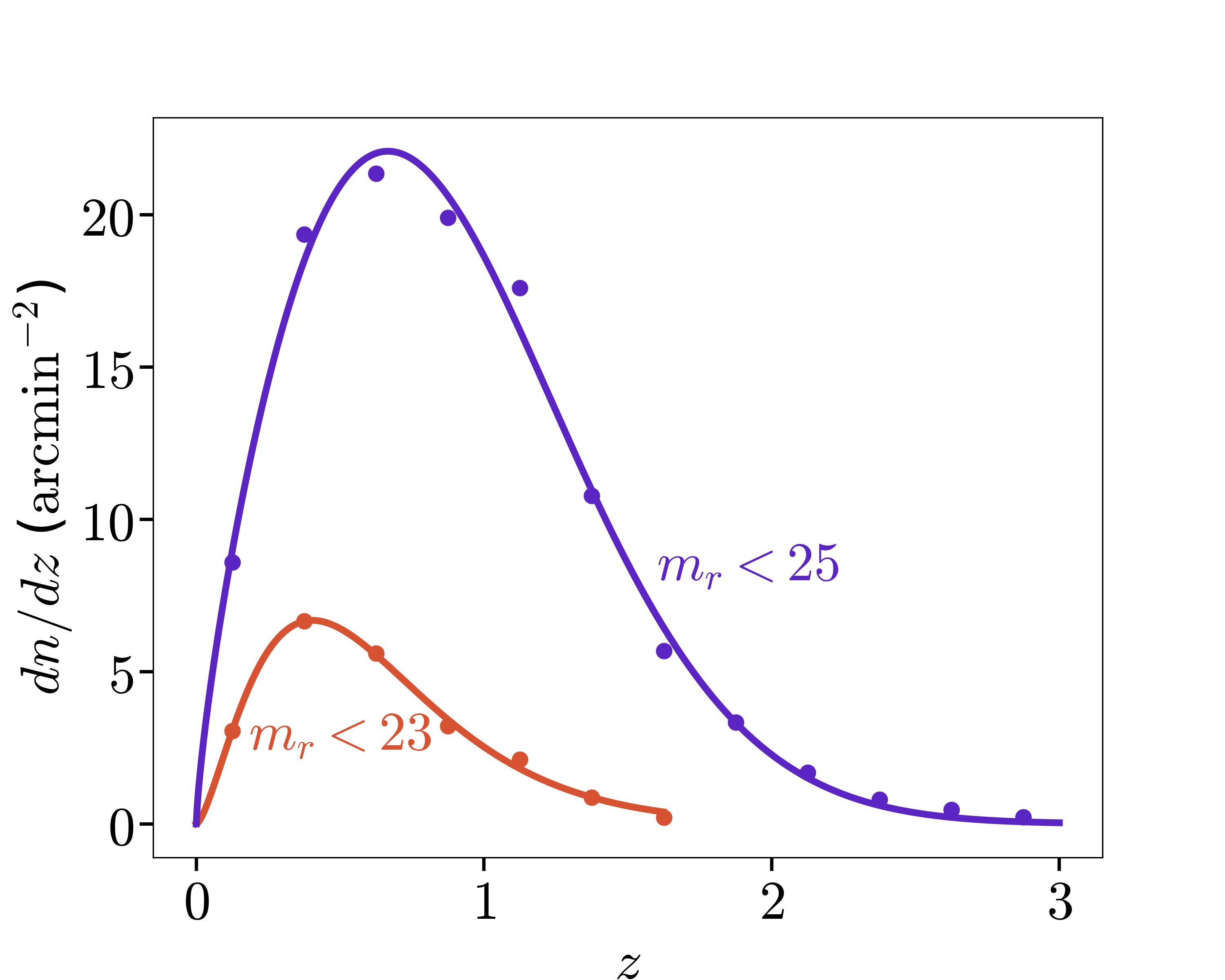}\label{fig:dndz}
        }
        \subfloat[\centering $b(z)$]{
        \includegraphics[width=0.33\textwidth]{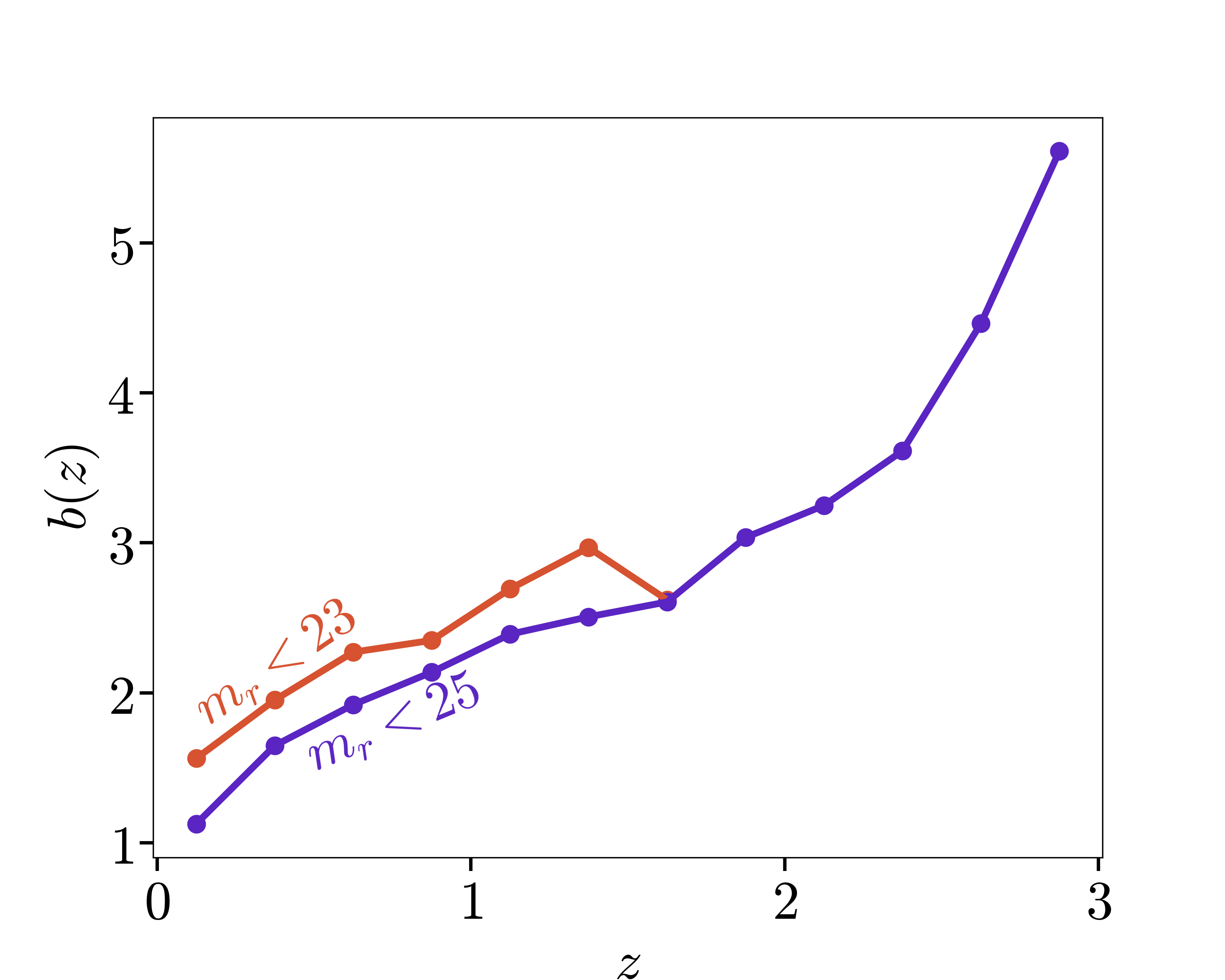}\label{fig:bias}
        }
        \subfloat[\centering $C_\ell$]{
        \includegraphics[width=0.33\textwidth]{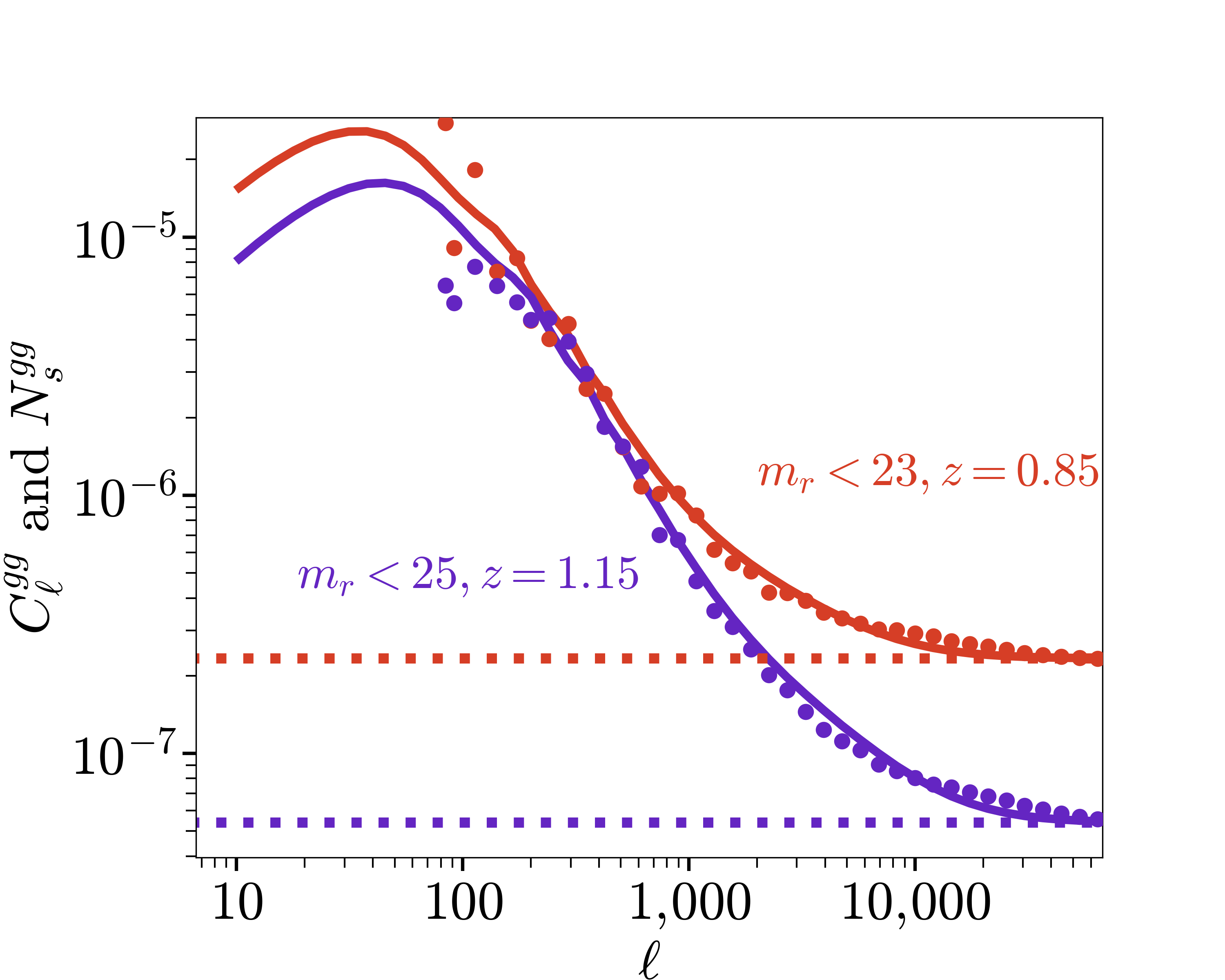}\label{fig:Cl}
        }
    \end{minipage}
    \caption{\textbf{(a)} $\dndz$ for the CosmoDC2 cutout. A fit was performed using the form in Eq. \eqref{eq:dndzFit} (solid line) to smooth the function for use in the calculation of $C_\ell$, $C_L$, and the SNR. The points used for performing the fit are shown as dots. \textbf{(b)} $b(z)$ for the CosmoDC2 cutout. \textbf{(c)} The galaxy angular power spectrum $C_\ell$ in $z$-bins of width $\Delta z=0.1$ for $m_r < 25$ and $m_r < 23$. The best-fit $C_\ell$ is represented by solid lines, while the shot noise floor $N_s=\frac{1}{\ngal(z)}$ corresponding to each $z$-bin and magnitude cut is indicated by the dotted lines. The $z$-bins shown here are the ones with the maximum \diffSNRLz.}
    \label{fig:ClComp}
\end{figure*}

We used a 25 \degsq cutout of CosmoDC2 \citep{CosmoDC2} to calculate the tomographic galaxy number density distribution \dndz and the linear galaxy bias $b(z)$ . We divided the CosmoDC2 cutout into 12 redshift bins of equal width with $0 < z < 3$ and calculated $n$ within each of these bins, which we used to estimate \dndz. To smooth the distribution, we parameterized \dndz as

\begin{align}
    \dndz=A\,\left(\frac{z}{z_0}\right)^m\,\exp\left[-\left(z/z_0\right)^\alpha\right]. \label{eq:dndzFit}
\end{align}
The form of this fit was chosen to be similar to \citep[p.~48]{LSSTReq}; however, because our redshift range was larger than theirs, we introduced the additional free paramter $m$. We found that for $m_r < 23$, $A=16.77\;\textnormal{arcmin}^{-2}$, $z_0=0.29$, $m=1.54$, and $\alpha=1.08$, while for $m_r < 25$, $A=46.12\;\textnormal{arcmin}^{-2}$, $z_0=1.09$, $m=0.75$, and $\alpha=2.04$.

The shot noise $N_s = \frac{1}{\ngal}$ within each $z$-bin was calculated by integrating \dndz (converting $\textnormal{arcmin}^{-2}$ to sr) over that $z$-bin; that is, for a $z$-bin with $z_0 < z < z_1$,
\begin{align}
    N_s(z) = \left(\int_{z_0}^{z_1}\dndz dz\right)^{-1}. \nonumber
\end{align}

The matter power spectrum $P_\delta\left(k, z\right)$ is modeled using \texttt{halofit} \citep{halofit}. Using \dndz and $P_\delta\left(k, z\right)$, we use Eq. \eqref{eq:Limber} setting
$b(z)=1$ to calculate the angular matter power spectrum.

To calculate $b(z)$, we first estimate the sample galaxy angular power spectrum $\hat{C}_\ell$ (which includes shot noise) directly from the CosmoDC2 cutout using the 12 $z$-bins we used to calculate \dndz. We then estimate the bias within each redshift bin by minimizing the residual of a weighted least-squares measure between the sample galaxy power spectrum $\hat{C}_\ell$ and the halofit power spectrum (plus noise) $C_\ell+N_s$:
\begin{align}
    &R(b)=\sum_\ell\frac{\left(\hat{C}_\ell(z) - \left[b^2 C_\ell(z)+N_s(z)\right]\right)^2}{\Delta C_\ell^2(z)}, \label{eq:LSqFit}\\
    &\Delta C_\ell^2(z)=\frac{2}{(2\ell+1)}\left(b^2 C_\ell(z)+N_s(z)\right)^2. \nonumber
\end{align}
Any required $b(z)$ for redshifts not exactly coinciding with the $z$-bins used for this calibration is obtained by linear interpolation.

Plots of \dndz, $b(z)$, and $C_\ell$ can be seen in Figure \ref{fig:ClComp}. For many CMB lensing reconstructions, a typical $\lmax$ in the CMB lensing field is around $3,\!000$. We will here use the galaxy number density field at full resolution, treating the shot noise as the limiting source of noise.

We will assume that the angular power spectrum for the LSST-like survey is measured over $20,\!000$ \degsq field ($f_\textnormal{sky}=0.48$), which matches the projected sky coverage of LSST \citep{LSSTHandbook}. For the DES-like survey, we will use a sky coverage of $5,\!000$ \degsq  \citep{DESGalLens}

\subsection{The theoretical convergence power spectrum}

The angular power spectrum for the convergence field at some comoving distance $\chi_s$ is given by \citep{SchaanCIB}

\begin{align}
    C_L(z)=\frac{9}{4}\frac{H_0^4}{c^4}\Omega_{m,0}^2\int_0^{\chi_s}\frac{\chi^2\left(\chi_s - \chi\right)^2}{a^2\chi_s^2}P_\delta\left(\frac{L}{\chi},z\right)\frac{d\chi}{\chi^2}. \label{eq:CL}
\end{align}

Lensing reconstruction using galaxy density fields as a source has great potential due to the massive amounts of galaxy position data available from existing surveys such as DES. This data has not been used for lensing reconstruction, so an efficient reconstruction method could be used to extract information from this otherwise untapped data source. This algorithm could also be used with new surveys such as LSST which provide unprecedented depth and precision in galaxy position measurements \citep{LSSTOverview,CatSim,CosmoDC2}. This information could be used to complement CMB lensing measurements in the same way that cosmic shear measurements can be used. This is an independent measurement {of the same physical effect as traditional galaxy shear measurements and is therefore a good complement to cosmic shear measurements}. This could in principle help improve the statistical power of observations and identify {possible systematic errors such as multiplicative errors in shear measurements. We could use the same galaxies for the shear measurements as for the lensing reconstruction, in which case the galaxy shape distortions and the distortion of the clustering should be a direct test for systematic errors.}

When calculating $C_L$ and $N_L$ for $m_r < 25$, we used all bins with $0 < z < 3$. However, for the DES-like model ($m_r < 23$), we performed these calculations only for $0 < z < 1.6$, as it was difficult to get
reliable clustering measurements at such low number density. Shear measurements from
DES Y3 extended only to redshift $z < 1$ \citep{DESGalLens}.

\section{Results}

In this section, we aim to determine the viability of using galaxy number density fields as a source for reconstructing lensing fields using either the standard quadratic estimator or a shear-only estimator. We will primarily focus on the LSST-like survey, then briefly discuss the equivalent calculations for a DES-like survey.

\subsection{Convergence power spectra}

\begin{figure}
    \centering
    \includegraphics[width=\textwidth]{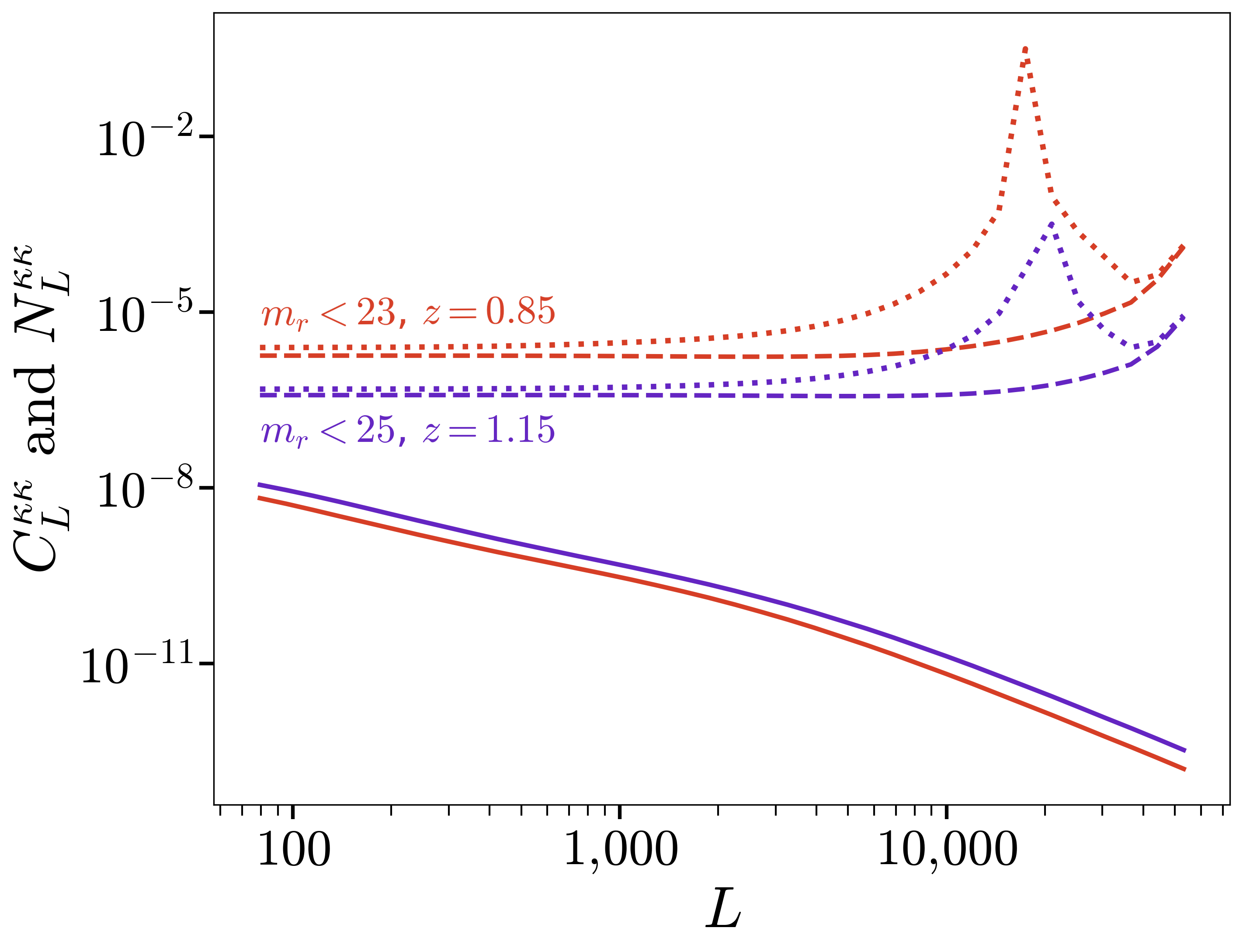}
    \caption{The convergence power spectra used for our reconstructions for $m_r < 23$ (orange line) and $m_r < 25$ (purple line). These consist of the convergence spectrum $C_L^{\kappa\kappa}$ (solid) and reconstruction noise $N_L$ (quadratic estimator: dashed, shear estimator: dotted). The $z$-bins used represent the bin with the maximum \diffSNRLz. The noise for shear estimator diverges at very large $L$, leading to the spike near $L=20,\!000$.}
    \label{fig:CLNL}
\end{figure}

The convergence spectrum $C_L^{\kappa \kappa}$ (solid lines) and reconstruction noise $N_L$ (quadratic estimator: dashed lines, shear estimator: dotted lines) are shown in Figure \ref{fig:CLNL} for two choices of magnitude cuts and $z$-bins with bin width $\dz=0.1$. This plot shows these quantities when integration was performed to $\lmax=70,\!000$. The noise is substantially higher than the signal at all $L$ for all choices of $z$, both for the traditional quadratic estimator and the shear estimator.

The reconstruction noise is roughly constant for small $L$; however, the integration kernels (Eq. \eqref{eq:Mkern} and \eqref{eq:Skern}) used to reconstruct the power spectra become less responsive to the signal at high $L$. This lack of response to signal manifests as a large spike in the noise as described in \citep{LensQuEst}. When integrating to much larger $\ell$ than is typically used for CMB lensing reconstructions \citep{LensQuEst}, the divergence of $N_L$ occurs at values of $L$ well beyond where the shape noise already dominates the cross-spectrum between the lensing reconstruction and cosmic shear measurements.

For verification, we calculate the total SNR of the autocorrelation function, defined as

\begin{align}
    \left(\frac{S}{N}\right)^2_a=\sum_{L,z}\frac{f_\textnormal{sky}(2L+1)}{2}\left(\frac{C_L^{\kappa\kappa}(z)}{C_L^{\kappa\kappa}(z)+N_L^{\kappa\kappa}(z)}\right)^2
\end{align}

This is equivalent to Eq. (4.1) in \citep{MBiasLensing}. For the survey parameters we use, the signal is much smaller than the noise for all $L$ as seen in Figure \ref{fig:CLNL}. Because of this, the SNR of the autocorrelation function will be small. We find that, for the LSST-like survey we use, the total SNR of the autocorrelation function for the standard quadratic estimator is $\left(\frac{S}{N}\right)_a=5$ when performing integration to $\ell=70,\!000$. {We sum over all $L$, but find that 75\% of the SNR comes from $L<100$ and 98.5\% comes from $L<3,\!000$.} {While small scales in the lensing field contribute little to the SNR of the autocorrelation spectrum because noise dominates this region,} including small scales from the source field has a large effect{, a result that is consistent with \citet{LuPen21a} and \citet{SchaanCIB}}. If we cut off our integration at $\ell=3,\!000$, the total SNR of the autocorrelation function for the standard quadratic estimator is only 0.5. This is also substantially (two orders of magnitude) smaller than the total signal to noise found in \citet{MBiasLensing}, as expected, where most of the signal to noise was coming from the inclusion of magnification bias.

\subsection{Signal-to-noise ratio for Cross-Spectra}

The lensing potential can be estimated in a wide variety of other ways. The most direct comparison would be with a measurement of the cosmic shear of galaxies at the same redshift as the source plane used for the reconstruction, but one could also estimate the expected lensing from adding up the contributions from the observed LSS tracers along the line of sight. 
We define the differential SNR for the cross spectrum between a different measurement of the lensing field, with total power spectrum $C_L+N_\gamma$ and the reconstructed signal, with power spectrum $C_L+N_L$, within an $L-z$-bin with widths \dz and $\Delta L$ as

\begin{align}
    \left(\frac{S}{N}\right)_\Delta(L, z)=\frac{C_L(z)}{\Delta C_L^{ij}(z)}, \label{eq:SNR}
\end{align}
where for our purposes \citep{SNREq}

\begin{align}
    &\left(\Delta C_L^{ij}\right)^2=\frac{1}{(2L+1)}\frac{\left(C_L^2+C_L^i C_L^j\right)}{f_\textnormal{sky}\,\Delta L}, \nonumber\\
    &\hat{C}_L^i=C_L+N_i, \label{eq:dCLx}\\
    &\hat{C}_L^j=C_L+N_j, \nonumber
\end{align}
where $C_L$ is the convergence power spectrum.

\begin{figure}
    \centering
    \centering
    \subfloat[\centering $m_r<25$, $\gamma_\textnormal{rms}=0$]{
    \includegraphics[width=\columnwidth]{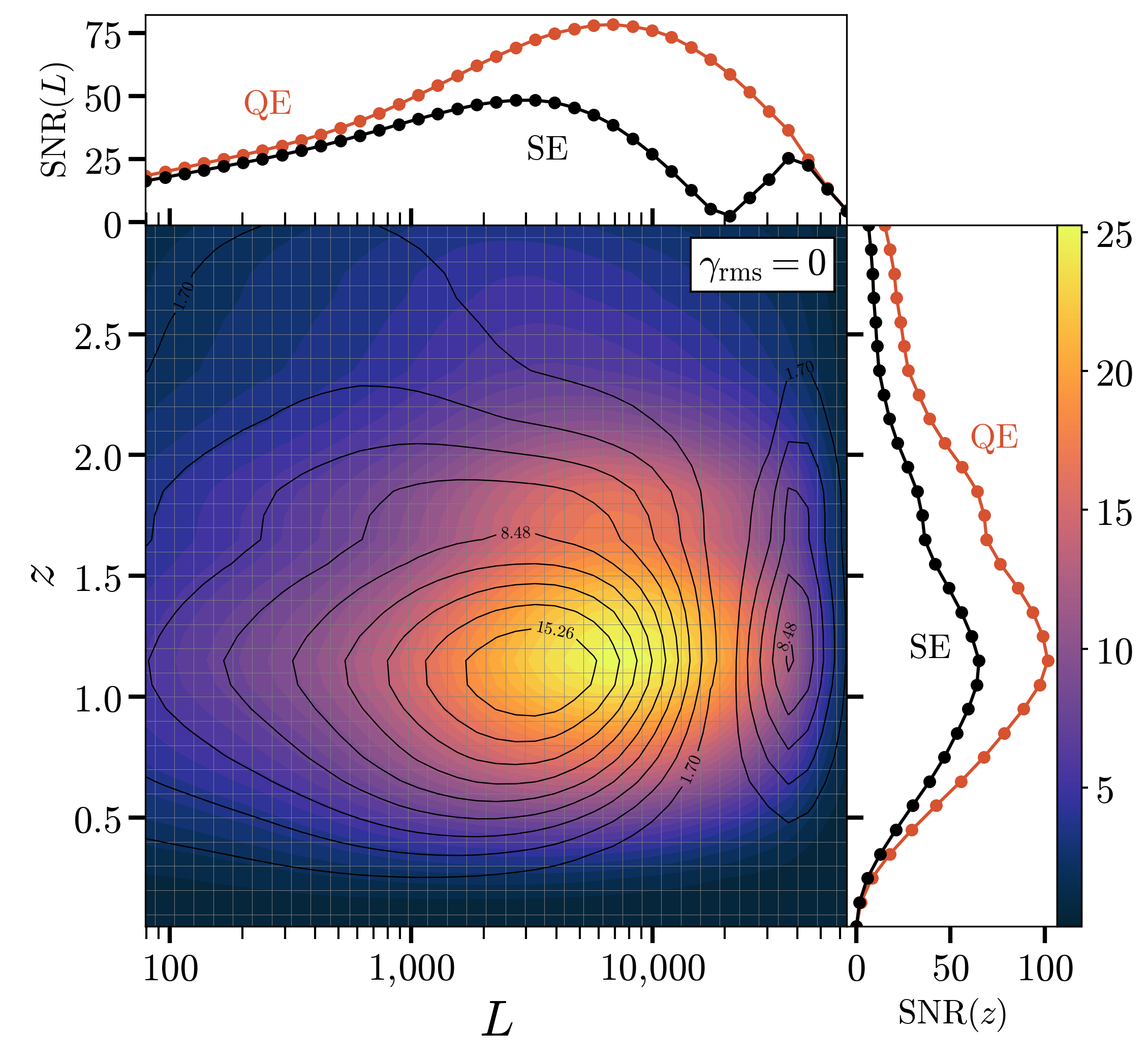}\label{fig:SNRnsn25}
    } \\
    \subfloat[\centering $m_r<25$, $\gamma_\textnormal{rms}=0.28$]{
    \includegraphics[width=\columnwidth]{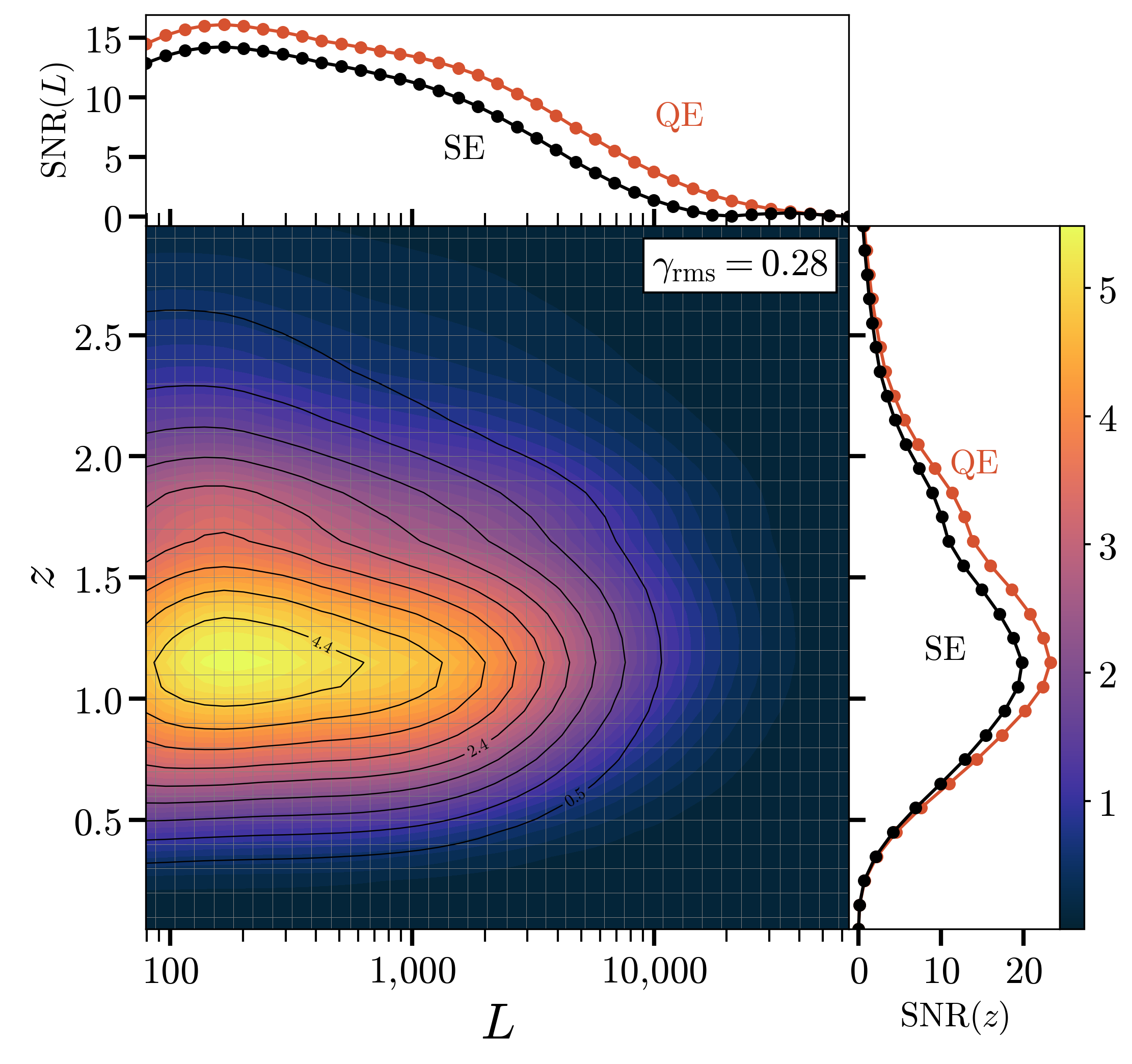}\label{fig:SNRsn25}
    }
    \caption{The differential SNR per $L-z$-bin (Eq. \eqref{eq:SNR}) for $\Delta\log L=0.08$, $\dz=0.1$ for the cross spectrum. The colored image shows the SNR for the standard quadratic estimator, while the black contours show the SNR for the shear estimator. The inset plots show the differential SNR per $L-z$-bin summed in quadrature over $z$ (top) and $L$ (right). We performed these calculations using 20,000 \degsq of the sky. In the upper plot, the sudden increase in the shear estimator SNR near $L=30,\!000$ corresponds with the region in $L$ beyond which the response of the shear estimator to the signal vanishes.}
    \label{fig:SNR25}
\end{figure}

The cross spectra \diffSNRLz (Eq. \eqref{eq:SNR}) for $m_r<25$ are shown in Figure \ref{fig:SNR25}. The colored background shows the quadratic estimator SNR, while the black contours show the shear estimator SNR.

Using Eqs. \eqref{eq:SNR} and \eqref{eq:dCLx}, we can obtain cumulative signal to noise for any collection of $z_j$ and $M_i$ bins by adding them in quadrature: $\textnormal{SNR}_\textnormal{cumulative}=\sum_{ij} \left(\frac{\textnormal{S}}{\textnormal{N}}\right)_\Delta(L_i, z_j)$.

Including all $L$ bins in the sum for a single $z$ bin leads to the total signal to noise for that $z$ bin. This is shown as the curve on the right side of Figure \ref{fig:SNR25}. Similarly, instead of summing over all $L$ bins for a given $z$ bin we can sum over all $z$ bins for a fixed $L$ bin to get the total signal to noise in a given $L$ bin, shown as the curve at the top of Figure \ref{fig:SNR25}. 

 Both without (top panel) and with (bottom panel) shape noise, \diffSNRLz peaked near $z=1.15$, reflecting the fact that the lensing signal is boosted by increased source distance while the noise amplitude increases at higher redshift as \dndz decreases, the latter of which falls off rapidly past $z \approx 1.3$.  Without shape noise, the shear estimator cross-spectrum \diffSNRLz is peaked at lower $L$ than the quadratic estimator, indicating that the shear estimator is less sensitive to information at smaller scales. The increase in the shear estimator SNR near $L=40,\!000$ represents the information that can be obtained beyond the noise spike near $L=10,\!000$. With shape noise, the $z$ and $L$ dependence of the quadratic estimator cross-spectrum is nearly identical to the shear estimator for \diffSNRLz, just with a larger magnitude.

\begin{figure*}
    \begin{minipage}{\textwidth}
    \centering
        \includegraphics[width=\textwidth]{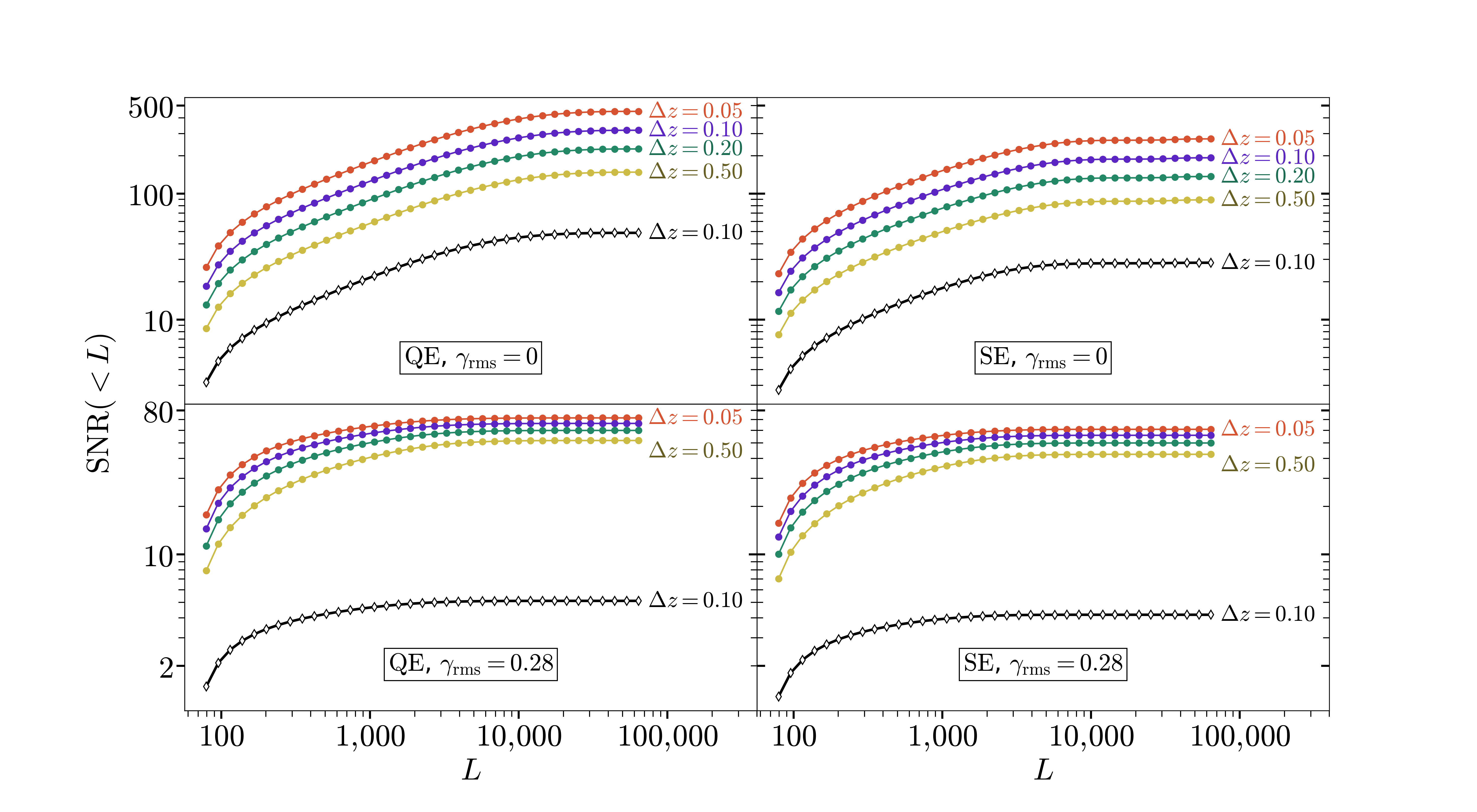}
    \end{minipage}
    \caption{Cumulative SNR in a cross-spectrum between a reconstruction and measurements of cosmic shear as a function of $L$, integrated over all redshifts, for (left) the standard quadratic estimator and (right) the shear estimator as a function of $L$ for various \dz values. Colored lines with circle markers represent the SNR for $m_r<25$, while the black line with white diamond markers shows the SNR for $m_r<23$. The top row shows the SNR without shape noise, while the bottom row shows it with LSST-like shape noise; the order of the \dz values from top to bottom remains the same. The total SNR increases as $\Delta z$ decreases.}
    \label{fig:SNRdz}
\end{figure*}

\subsection{\dz dependence}

When performing tomographic measurements, it is important to ensure that each $z$-bin contains a sufficient number of galaxies to prevent shot noise from dominating the signal. However, making \dz as small as possible will generally improve reconstruction accuracy because $C_\ell$ will vary less within the bin as well as having simply more measurements of the line of sight lensing potential. While a wide $z$-bin will generally contain more galaxies (hence lowering the shot noise and shape noise), $C_\ell$ will also be smaller due to projection effects from the larger integration range in $z$. Unfortunately, depending on the magnitude cut, survey depth, photo-$z$ uncertainties, and other factors, the minimum width of $z$-bins will be limited. Thus, we will assess how our results change when varying \dz.

The cumulative SNR for all $z$-bins combined for an LSST-like survey for various \dz values is shown in Figure \ref{fig:SNRdz} (colored lines with circle markers). This SNR was calculated by subdividing the full $0 < z < 3$ range into bins with four different widths. For a given $L-z$-bin, neither $C_L$ nor $N_L$ vary with \dz because the effects of galaxy number counts on shot noise are exactly canceled by the projection effects from performing the integration in Eqs. \eqref{eq:Limber} and \eqref{eq:CL}. From Eq. \eqref{eq:SNR}, and using the fact that the reconstructions are always strongly noise-dominated, it can be seen that the signal to noise in a single $L,z$ bin will be unchanged by the size of the $z$ bin in the absence of shape noise.
Because smaller bins allow for more bins in redshift, the integrated SNR will increase as  $1/\sqrt{\dz}$ in the absence of shape noise. 

However, the shape noise $N_\gamma$ does change with \dz because it depends on the number of galaxies within a given $z$-bin. When we include shape noise, the SNR even for a single $L,z$ bin does vary with bin width. At small $L$, the signal in the shear-based reconstruction still dominates over shape noise, and at high $L$, the shape noise overtakes the signal at larger $L$ and the cumulative SNR as a function of $L$ flattens out for all \dz.

\subsection{DES-like survey}

The Dark Energy Survey already provides a massive amount of galaxy data that could currently be used for lens reconstruction. To determine the effectiveness of our method using this dataset, we replicated the above calculations for a DES-like survey. We assumed a $5,\!000$ $\textnormal{deg}^2$ DES-like survey with a maximum $r$-band magnitude of 23 and used a redshift range of $0 < z < 1.6$ \citep{DESGalLens}. While the SNR can theoretically be calculated for higher $z$, our method for calculating $b(z)$ was not valid for $z \gtrapprox 1.6$ because the shot noise in the mock catalogs completely dominated the signal. Early estimates for the rms shape noise (not including measurement error) for a DES-like survey were $\gamma_\textnormal{rms}=0.25$ \citep{DESTaskForce}, while the standard shear noise estimate (which includes both shape noise and measurement error $\sigma_m$) is $\gamma_\textnormal{rms}^2+\sigma_m^2=0.255^2$ \citep{Chang13}; as a conservative estimate and to match the parameters of the LSST-like survey, we will estimate the rms shear noise as $\gamma_\textnormal{rms}=0.28$ when calculating the SNR.

Figure \ref{fig:CLNL} shows $C_L$ and $N_L$ for the full $5,\!000$ $\textnormal{deg}^2$ DES-like survey for the $z$-bin that maximized \diffSNRLz. The increased shot noise increased the reconstruction noise $N_L$ by as much as an order of magnitude for some $z$-bins relative to the LSST-like survey. There, we again see that the signal is much smaller than the noise for all $L$.

Figure \ref{fig:SNRdz} shows the full-survey SNR for $m_r<23$, $\dz=0.1$ (black line with white diamond markers). While the total SNR with shape noise is substantially lower than for the LSST-like survey (about 5.6 for the quadratic estimator, 4.6 for the shear estimator for $\dz=0.1$) due to the increased $N_L$, it remains above unity. This indicates that it may be possible to use DES galaxy number density data as a source field for lensing reconstruction; however, a more thorough analysis of additional error contributions is required. In addition, we found that there was substantial signal for $z>1.6$, indicating that more information could be obtained from a DES-like survey if the bias calculations could be performed at higher $z$.

\section{Conclusion}

We demonstrated that a galaxy number density field can be used as a source for convergence field reconstruction. We performed our reconstructions with a modification of the standard quadratic estimator and a shear-only estimator using a background galaxy density field as a source distribution. We demonstrated that this method can be used to reconstruct the convergence power spectrum $C_L$ and we estimated the SNR in cross-correlation that can be obtained from an LSST-like survey within various tomographic bins at two magnitude cuts, ranging from barely detectable with $m_r<23$ to over 50 for $m_r<25$.

The total SNR using both our standard quadratic estimator and shear estimator increases with galaxy number density, indicating that increasing the depth and sensitivity of surveys will improve the SNR within each tomographic bin. We expect that using the full LSST 5$\sigma$ 10 year coadd. dataset would increase the SNR. However, we still expect to get a high SNR when using only single snapshots from an LSST-like survey. The SNR in cross-correlation continues to increase as the redshift resolution increases, indicating that accurate photometric redshifts will be of great value.

Whether this reconstruction method will be useful for DES data is unclear. The total SNR is greater than unity for both estimators, but the inclusion of additional error sources will decrease the SNR. The information that we obtained in our forecast from the DES-like survey is also somewhat limited by the $z=1.6$ cutoff imposed in our analysis. If the galaxy clustering can be measured at higher redshift, we expect that more information might be obtained. Future work should recalculate the SNR for a DES-like survey beyond this $z=1.6$ cutoff.

The analysis we performed used only reconstruction noise and shape noise as sources of error for $C_L$; it did not include some error sources associated with survey measurements. These include foreground contamination and photometric redshift and other measurement uncertainties. Future work should recalculate the SNR with these error sources included. This would provide information about the effectiveness of our method for performing convergence field reconstruction from observed galaxy survey data. 

{For this forecast, we calculate the galaxy power spectrum using \texttt{halofit} and the linear bias, which are not accurate at very small scales. In practice, the galaxy power spectrum will be already well-measured before performing the lensing reconstruction. This introduces uncertainty in our forecast, but uncertainty in the galaxy power spectrum will not be a limiting factor for data analysis. This uncertainty in the accuracy of the non-linear matter power spectrum calculation will be important for interpreting the reconstructed convergence spectra because some of the SNR comes from small scales, especially in the limit of a low-noise tracer of the density field being used for cross-correlation.}

A substantial concern that we have neglected for this analysis is the non-Gaussianity of the galaxy number density counts on small scales, exactly the scales where we are obtaining most of our lensing information.
Correcting for non-Gaussianities would help further improve our estimator. The non-Gaussianities and strong nonlinearities at small scales in galaxy density fields may make higher-order moments of the galaxy density field (the bispectrum, trispectrum, etc.) non-negligible and introduce additional noise. Line intensity mapping (LIM) source fields are highly nonlinear because they, like the galaxy number counts, trace the underlying matter distribution. Reconstruction methods designed for 21 cm LIM \citep{Pen21,ZahnZal21,ForemanLIM,LuPen21b} may be applicable for reconstruction using a galaxy density field as a source. Future work should evaluate the impact of non-Gaussianities and foreground contamination on the SNR of reconstructions using our estimator. We would evaluate if, like the estimators described by \citet{LuPen21a} and \citet{SchaanCIB}, the shear estimator is less susceptible to foreground contamination than the standard quadratic estimator when using a galaxy density source field. The impact of non-Gaussianities should be compared with reconstructions using the methods of \citet{ForemanLIM} and \citet{LuPen21b} for a galaxy density field background because these estimators are designed to reduce non-Gaussian noise (or, for the case of \citet{LuPen21b}, are optimal for non-Gaussian fields). {The impact of non-Gaussianities, as well as the robustness of the QE when performing $\ell$-integration to scales much smaller than are typically used for CMB lensing reconstruction, ultimately must be evaluated by performing lensing reconstructions on mock catalogs, a task left for future work.}

\section{Software Used}

Our calculations of the angular power spectra, convergence spectra, and reconstruction noise were performed using code based off of \texttt{LensQuEst} \citep{LensQuEst}. CosmoDC2 data \citep{CosmoDC2} was used for various calculations and was obtained using {\tt GCRCatalogs} \citep{GCRCat} and {\tt CatSim} \citep{CatSim}.

\section{Acknowledgements}

 This work was supported by the Brand \& Monica Fortner Chair in Theoretical Astrophysics
and by the National Science Foundation Graduate Research Fellowship Program Grant No. DGE --- 1746047.

This work utilizes resources supported by the National Science Foundation’s Major Research Instrumentation program, grant \#1725729, as well as the University of Illinois at Urbana-Champaign \citep{HAL}.

This research was supported in part by Perimeter Institute for Theoretical Physics. Research at Perimeter Institute is supported by the Government of Canada through the Department of Innovation, Science and Economic Development Canada and by the Province of Ontario through the Ministry of Research, Innovation and Science.

\bibliographystyle{aasjournal}
\bibliography{main}

\label{lastpage}
\end{document}